\documentclass[9pt, conference]{IEEEtran}
\IEEEoverridecommandlockouts
\usepackage{cite}
\usepackage{amsmath,amssymb,amsfonts}
\usepackage{algorithmic, algorithm}
\usepackage{graphicx}
\usepackage{textcomp}
\usepackage{xcolor}
\usepackage{comment}
\usepackage{bm}
\usepackage{xspace}
\usepackage{enumitem}

\renewcommand{\baselinestretch}{0.94}

\def\BibTeX{{\rm B\kern-.05em{\sc i\kern-.025em b}\kern-.08em
    T\kern-.1667em\lower.7ex\hbox{E}\kern-.125emX}}
\begin{document}

\vspace{-3mm}
\title{\huge{LinEasyBO: Scalable Bayesian Optimization Approach for Analog Circuit Synthesis via One-Dimensional Subspaces}\\
\author{Shuhan Zhang$^{1*}$, Fan Yang$^{1*}$, Changhao Yan$^1$, Dian Zhou$^2$, Xuan Zeng$^{1*}$\\
$^1$State Key Lab of ASIC \& System, Microeletronics Department, Fudan University, China\\
$^2$University of Texas at Dallas, Dallas, USA\vspace{-5mm}}
\vspace{-3mm}
\thanks{*Corresponding authors: \{shuhanzhang16, yangfan, xzeng\}@fudan.edu.cn}
}

\begin{comment}
\author{\IEEEauthorblockN{1\textsuperscript{st} Given Name Surname}
\IEEEauthorblockA{\textit{dept. name of organization (of Aff.)} \\
\textit{name of organization (of Aff.)}\\
City, Country \\
email address or ORCID}
\and
\IEEEauthorblockN{2\textsuperscript{nd} Given Name Surname}
\IEEEauthorblockA{\textit{dept. name of organization (of Aff.)} \\
\textit{name of organization (of Aff.)}\\
City, Country \\
email address or ORCID}
\and
\IEEEauthorblockN{3\textsuperscript{rd} Given Name Surname}
\IEEEauthorblockA{\textit{dept. name of organization (of Aff.)} \\
\textit{name of organization (of Aff.)}\\
City, Country \\
email address or ORCID}
\and
\IEEEauthorblockN{4\textsuperscript{th} Given Name Surname}
\IEEEauthorblockA{\textit{dept. name of organization (of Aff.)} \\
\textit{name of organization (of Aff.)}\\
City, Country \\
email address or ORCID}
\and
\IEEEauthorblockN{5\textsuperscript{th} Given Name Surname}
\IEEEauthorblockA{\textit{dept. name of organization (of Aff.)} \\
\textit{name of organization (of Aff.)}\\
City, Country \\
email address or ORCID}
\and
\IEEEauthorblockN{6\textsuperscript{th} Given Name Surname}
\IEEEauthorblockA{\textit{dept. name of organization (of Aff.)} \\
\textit{name of organization (of Aff.)}\\
City, Country \\
email address or ORCID}
}
\end{comment}

\maketitle

\begin{abstract}
A large body of literature has proved that the Bayesian optimization framework is especially efficient and effective in analog circuit synthesis. However, most of the previous research works only focus on designing informative surrogate models or efficient acquisition functions. Even if searching for the global optimum over the acquisition function surface is itself a difficult task, it has been largely ignored. In this paper, we propose a fast and robust Bayesian optimization approach via one-dimensional subspaces for analog circuit synthesis. By solely focusing on optimizing one-dimension subspaces at each iteration, we greatly reduce the computational overhead of the Bayesian optimization framework while safely maximizing the acquisition function. By combining the benefits of different dimension selection strategies, we adaptively balancing between searching globally and locally. By leveraging the batch Bayesian optimization framework, we further accelerate the optimization procedure by making full use of the hardware resources. Experimental results quantitatively show that our proposed algorithm can accelerate the optimization procedure by up to $9\times$ and $38\times$ compared to LP-EI and REMBOpBO respectively when the batch size is 15.
\end{abstract}

% \begin{IEEEkeywords}
    
% \end{IEEEkeywords}

\section{Introduction}

% As the development of the integrated circuit (IC) technology, the circuit design has become more and more complicated. With the increasing demands of low power, high performance, short time-to-market, designing circuits manually has become increasingly difficult and even impossible. Although the digital circuit design has been automated for a long time, the analog circuit design automation still has a long way to go. Sophisticated analog circuit design automation tools are in great need from both industry and academia. In general, the analog circuit design process can be divided into two stages: the topology selection and device sizing. In this paper, we mainly focus on the device sizing problem.

As the development of the integrated circuit (IC) technology, the circuit scale increases dramatically and the unwanted parasitic effect complicates the circuit design process significantly. The increasing demands on speed, reliability, low power consumption and short time-to-market further aggravate the difficulty of designing circuit manually, especially for the analog circuit \cite{rutenbar2007hierarchical}. Sophisticated analog circuit design automation tools are in great need. The analog circuit design procedure generally can be divided into two stages: topology selection and device sizing. In this paper, we mainly focus on the device sizing problem.

% With the increasing demands on high-speed, high-reliability, low-power, and short-to-mark

% the unwanted parasitics effects and the increasing scale of the circuits has made the circuit design a much more complicated task. With the increasing demands on high-speed, low-power, high-reliability and short-to-market, designing circuits manually has been increasingly difficult and even intractable sometimes. 

% the unwanted effect of circuit parasitics has larger impact on the circuit performance. 

% The analog circuit design process can be divided into two stages: the topology selection and the device sizing. In this paper, we mainly focus on the device sizing problem.

% The analog circuit device sizing problem typically can be formulated into a noisy, multimodal, expensive to evalu

The analog circuit device sizing problem typically can be formulated as a noisy, multimodal, expensive and derivative-free black-box function. A large body of literature has been published on designing an efficient optimization algorithm to solve this problem \cite{liu2009analog,liu2014gaspad:,phelps2000anaconda:,rutenbar2007hierarchical,hu2018parallelizable,yang2018smart-msp:,lyu2017efficient}. The traditional optimization algorithms for analog circuit device sizing problem can be divided into two groups: the model-based and simulation-based approaches. The model-based approaches try to approximate the circuit performance and provide a cheap-to-evaluate model for space exploration. Instead of searching for the global optimum with expensive circuit simulator, the constructed model can greatly reduce time consumption on observation, thus, accelerate the optimization procedure. One popular model-based approach is the geometric programming \cite{boyd2005geometric, boyd2007tutorial, hannah2012ensemble}. However, there is no theoretical guarantee that the constructed model is accurate over the whole design space. Therefore, the obtained global optimum may deviate from the real global optimum.

Instead of constructing a surrogate model to capture the behavior pattern of the circuit performance, the simulation-based approaches simply search the state space by mimicking the physical or biological phenomenon. By treating the circuit performance as a black-box function, the simulation-based approaches guide the search with a selection engine and invoke the circuit simulator on the fly. There are several popular simulation-based approaches, including simulated annealing (SA) \cite{gielen1990analog, grzechca2007use}, multiple starting point (MSP) algorithm \cite{yang2018smart-msp:, peng2016efficient}, evolutionary algorithm \cite{phelps2000anaconda:, alpaydin2003evolutionary} and particle swarm optimization (PSO) algorithm \cite{fakhfakh2010analog, wu2009a}. The limitation of the simulation-based approaches is their relatively low convergence rate.

Inspired from above, the Bayesian optimization framework has been proposed to pave a way out of the dilemma by combining the benefits of both model-based and simulation-based methods \cite{shahriari2015taking, brochu2010tutorial}. Generally, there are two key elements in the Bayesian optimization framework: the statistical surrogate model and the acquisition function. Injected with our prior beliefs about the latent function, the surrogate model provides both the predictive mean and uncertainty estimation for each location based on the collected data points. By leveraging the posterior distribution provided by the surrogate model, the acquisition function works as a utility function to measure the potential of each location over the design space and observe the most promising data points in an iterative and greedy manner. After a limited number of iterations, the global optimum can be reached with a theoretical guarantee \cite{srinivas2009gaussian, wang2017max}. Compared to the model-based methods that only build the model once and searches for the global optimum offline, the Bayesian optimization approach incrementally updates the surrogate model and iteratively invokes the circuit simulator on the fly. Compared to the simulation-based methods that solely explores the state space with a heuristically designed selection engine, the Bayesian optimization approach makes full use of the previously collected information and facilitates the searching procedure more economically. Prior researches have demonstrated the efficiency and effectiveness of the Bayesian optimization framework for analog circuit synthesis \cite{liu2009analog,liu2014gaspad:,liu2016multi,lyu2017efficient}.

Although lots of efforts have spent on designing an informative surrogate model and efficient acquisition function \cite{zhang2019bayesian, zhang2019efficient, zhang2020efficient,liu2014gaspad:, lyu2017efficient, lyu2018batch, srinivas2009gaussian, hu2018parallelizable, he2020efficient}, the acquisition function optimization procedure itself is always assumed to be relatively easy and largely ignored. However, the global optimum can only be theoretically guaranteed for the Bayesian optimization framework when it successfully finds the global optimum of the acquisition function at each iteration. The efficiency of the Bayesian optimization framework can be significantly hurt if less promising data points are accumulated across iterations. In other words, the acquisition function optimization procedure is important both theoretically and practically. With the increase of the dimensionality, this problem only deteriorates and further sabotages the efficiency of the Bayesian optimization framework. In this paper, we propose a fast and robust Bayesian optimization approach via one-dimensional subspaces \cite{kirschner2019adaptive} for analog circuit synthesis. Unlike the traditional Bayesian optimization methods that optimize the acquisition function over the entire design space, we solely focus on one dimension at a time. In this way, our proposed algorithm makes solid progress at each iteration and ensures information gain regardless of the dimensionality. We also leverage the carefully designed batch Bayesian optimization framework to further increase sampling efficiency. Experimental results quantitatively demonstrate that our proposed algorithm can accelerate the optimization procedure by up to $9\times$ and $38\times$ compared to LP-EI \cite{gonzalez2016batch} and REMBOpBO \cite{hu2019enabling} respectively while obtaining the same optimization results when the batch size is 15.

The rest of the paper is organized as follows. In \S\ref{sec:background}, we formulate the device sizing problem and briefly review the knowledge background of our proposed algorithm. In \S\ref{sec:proposal}, we analytically highlight the challenges of the acquisition function optimization procedure and our proposed method of handling these problems. In \S\ref{sec:experiments}, we conduct experiments on two real-world analog circuits and quantitatively demonstrate the efficiency and effectiveness of our proposed algorithm. 
Finally, we conclude the paper in \S\ref{sec:conclusion}.

\section{Background} \label{sec:background}

In this section, we present both the problem formulation of device sizing (\S\ref{sec:problem_formulation}) and a brief review of the Bayesian optimization framework (\S\ref{sec:bayesian_optimization}).

\subsection{Problem Formulation} \label{sec:problem_formulation}

Given a topology selection, the analog circuit device sizing problem can be formulated into a black-box function optimization problem, which can be expressed as follow:
\begin{equation}
    \label{eq:problem_formulation}
    \text{minimize} \quad f(\bm{x}).
\end{equation}
By searching the $d$-dimensional design space, our target is to find an optimal circuit design $\bm{x}$ that minimizes the Figure of Merit (FOM) $f(\cdot)$ over the design space $\mathcal{X}$. For each design $\bm{x}$, we regard the simulation results generated by circuit simulators including commercial HSPICE and Spectre software as the ground-truth circuit performances $f(\bm{x})$.

% the simulation results generated by circuit simulators including the commercial HPSICE and Spectre software are

% the corresponding ground-truth value of circuit performance $f(\bm{x})$ is generated by circuit simulators including the commercial HPSICE or Spectre circuit simulators.

\subsection{Review of the Bayesian Optimization Framework} \label{sec:bayesian_optimization}

There are two basic elements in the Bayesian optimization framework: the surrogate model and the acquisition function \cite{shahriari2015taking}. The surrogate model captures our prior beliefs about the latent function and provides an informative posterior distribution for each location over the design space. The posterior distribution means both the predictive mean and uncertainty estimation are provided. The predictive mean describes the behavior pattern of the latent function predicted by the surrogate model. The uncertainty estimation shows how much confidence the model has on its predictive mean. In other words, the surrogate model not only tells us what it believes but also how much we can trust its beliefs. Compared to deterministic models that only provide predictive mean and guide the search in a greedy manner, the stochastic statistical surrogate model employed by the Bayesian optimization framework can better facilitate the searching process and help to make more rational decisions. 

One commonly used surrogate model is the Gaussian process regression (GPR) model. Assume the we have an initial dataset $D=\{X, \bm{y}\}$, where $X=\{\bm{x}_1, \cdots, \bm{x}_N\}$,  $\bm{y}=\{y_1, \cdots, y_N\}$, and $N$ is the size of the dataset. By injecting our prior assumption about the latent function with mean function $m(\cdot)$ and kernel function $k(\cdot, \cdot)$, the posterior distribution for each location $\bm{x}_\ast$ can be expressed as \cite{rasmussen2003gaussian}:
\begin{equation}
    \label{eq:prediction}
    \begin{cases}
        \begin{aligned}
        \mu(\bm{x}_\ast) & = m(\bm{x}_\ast) + k(\bm{x}_\ast) [K+\sigma_n^2 I]^{-1}(\bm{y} - \bm{m})\\
        \sigma^2(\bm{x}_\ast) & = k(\bm{x}_\ast, \bm{x}_\ast) - k(\bm{x}_\ast, X)[K+\sigma^2_n I]^{-1}k(X, \bm{x}_\ast),
        \end{aligned}
    \end{cases}
\end{equation}
where $\bm{m} = (m(\bm{x}_1), \cdots, m(\bm{x}_N))^T$, $K=(k(\bm{x}_1, X), \cdots, k(\bm{x}_N, X))^T$, and $\sigma^2_n$ is the noise variance. In this paper, we fix the mean function as $m(\bm{x}) = 0$ and the kernel function as the square exponential (SE) covariance function.

By taking both the predictive mean and uncertainty estimation into consideration, the acquisition function works as a utility function to prioritize each location over the design space. By favoring the potential area with high uncertainty estimation, the acquisition function explores the unknown region to avoid getting stuck in the local optimum. By selecting locations with high probability to be optimal, the acquisition function exploits the collected information to avoid too much sampling around the less promising area. In other words, the acquisition function trades off between searching globally (exploration) and locally (exploitation). Popular acquisition functions include the probability of improvement (PI) \cite{kushner1964new}, expected improvement (EI) \cite{mockus1978application}, lower confidence bound (LCB) \cite{srinivas2009gaussian}, Thompson sampling (TS) \cite{thompson1933likelihood} and entropy search (ES) \cite{hennig2012entropy}. A portfolio of several acquisition functions is also possible \cite{hoffman2011portfolio, lyu2018batch}.

The Bayesian optimization framework starts the optimization procedure by first randomly sampling an initial dataset. By embedding our prior beliefs about the unknown objective function, the surrogate model builds upon the collected data points and provides the posterior distribution for each location over the design space. By taking both exploration and exploitation into consideration, the acquisition function helps to select the query points and traverse the design space efficiently. At each iteration, selected data points are observed to incrementally update the dataset. With a limited number of iterations, the Bayesian optimization framework can theoretically obtain the global optimum. The corresponding framework of the Bayesian optimization algorithm is presented in Algorithm \ref{algo:bayesian_optimization}.

% By first randomly sampling an initial dataset, the surrogate model helps to provide the posterior distribution for each location in the design space. By taking both predictive mean and the uncertainty estimation into consideration, the acquisition function helps to select data points that achieve a better tradeoff between the exploration and exploitation. After evaluating the selected data points and adding it to the dataset, the surrogate model can be updated to provide more informative posterior distribution. With a limited iterations, the Bayesian optimization framework can theoretically obtain the global optimum. The corresponding framework for the Bayesian optimization approach is presented in Algorithm \ref{algo:bayesian_optimization}.

\begin{algorithm}
    \caption{Bayesian Optimization Framework}
    \label{algo:bayesian_optimization}
    \begin{algorithmic}
        \STATE Randomly sample an initial dataset $D_0=\{X, \bm{y}\}$
        \FOR{$t = 1$ to $T$}
        \STATE Build a Gaussian process regression model with $D_{t-1}$
        \STATE Select $\bm{x}_t$ that optimizes the acquisition function
        \STATE Observe the selected data $y_t = f(\bm{x}_t)$
        \STATE Update the dataset $D_t = \{D_{t-1}, \{\bm{x}_t, y_t\}\}$
        \ENDFOR
        \RETURN The best $f(\bm{x})$ recorded during optimization
    \end{algorithmic}
\end{algorithm}

% \subsection{High-dimensional Bayesian Optimization}

% Large quantity of research works have demonstrated the efficiency and effectiveness of the Bayesian optimization framework. However, due to the curse of dimensionality, the performance of the Bayesian optimization is greatly limited to a relatively low-dimensional design space. In order to further extend the application of the Bayesian optimization framework, a large body of literature has been published to solve this problem.

\section{Proposed Approach} \label{sec:proposal}

In this section, we first highlight the fundamental challenges of the acquisition function optimization procedure and how it affects the efficiency of the Bayesian optimization framework (\S\ref{sec:challenges}). Then, we explicitly describe our propose algorithm to handle this problem (\S\ref{sec:linebo} and \S\ref{sec:dimension_selection}). Finally, we summarize the proposed method (\S\ref{sec:summary}).

\subsection{Challenges of Acquisition Function Optimization} \label{sec:challenges}

% The success of the Bayesian optimization framework relies heavily on a series of basic elements, such as the surrogate model and the acquisition function. The efficiency and effectiveness of these elements has been 

A large body of literature has been published to improve the efficiency of the Bayesian optimization framework. However, most of the previous researches mainly focus on designing a more informative surrogate model or an efficient acquisition function. The inner optimization procedure (\textit{a.k.a} the acquisition function optimization procedure) has been assumed to relatively easy and often neglected. 

% The acquisition function optimization procedure is also called the inner optimization procedure.

% theoretically important
From a \textit{theoretical} point of view, the global optimum is only granted with a limited number of iterations for the Bayesian optimization framework, if the global optimum of the acquisition function is obtained at each iteration. In other words, the success of the Bayesian optimization framework crucially relies on successfully optimizing the acquisition function. The less promising data points accumulated across iterations can violate the theoretical justification of the Bayesian optimization framework. Therefore, it is theoretically important to optimize the acquisition function optimally.

% practically difficult
From a \textit{practical} point of view, the efficiency of the acquisition function optimization procedure suffers greatly from the \textit{curse of dimensionality}. With the increase of the dimensionality, the state space for exploration increases exponentially. To search for the global optimum globally and make full coverage of the design space, the required number of acquisition function estimations can be prohibitively large. Take the grid search as an example, let us assume that we evenly select $M$ data points for acquisition function estimation at each dimension to make full coverage of the design space. The expected number of data points for acquisition function estimation is $M^d$ when the design space is $d$-dimensional. Considering that the computational complexity of the training and prediction procedure for GPR model are $O(N^3)$ and $O(N^2)$ respectively, the computational complexity for acquisition function optimization is $O(N^2M^d)$. Therefore, even if a large body of literature claims that the theoretical computational bottleneck of the Bayesian optimization framework is the model training process, the acquisition function optimization procedure easily dominants the computational overhead in practice. Although other algorithms have also been studied to facilitate the acquisition function optimization procedure including MSP algorithm \cite{lyu2017efficient}, stochastic gradient ascent algorithm \cite{wilson2018maximizing}, evolutionary strategy \cite{lyu2018multi}, DIRECT \cite{jones1993lipschitzian}, etc. This problem is only alleviated not solved. 

In summary, the acquisition function optimization procedure is theoretically important and practically difficult. We tackle this problem by proposing a scalable Bayesian optimization framework in this paper.

\subsection{Optimization in One-Dimensional Subspaces} \label{sec:linebo}

% With the curse of dimensionality, the acquisition function optimization procedure becomes practically intractable when the dimension of the design space is relatively large. 
Although optimizing the acquisition function over the whole design space is practically expensive or even intractable, searching for the global optimum of the acquisition function over the one-dimensional subspace at a time is relatively easy \cite{scarlett2018tight}. Instead of optimizing the acquisition function all over the design space, we constrain the searching space to a one-dimensional subspace $\mathcal{L}_i = \{\bm{x}_\ast + \beta l_i : \beta \in \mathbb{R}\} \cap \mathcal{X}$ that contains the previously sampled best point $\bm{x}_\ast$ \cite{kirschner2019adaptive}, where $\mathcal{L}_i$ represents the $i$-th one-dimensional subspace and $l_i$ denotes the $i$-th direction. In this way, the global optimum of the acquisition function can be effectively obtained by grid search. Without surrendering the efficiency of the acquisition function optimization procedure, our proposed algorithm makes solid progress and ensures information gain at each iteration. 

With the increase of the dimensionality, the traditional way of handling the acquisition function optimization procedure tends to follow the same behavior pattern as random search. Our proposed method instead makes solid and safe progress and maximizes information gain in the selected one-dimensional subspace, thus, improves the sampling efficiency.

% To chart a way out of the curse of dimensionality, we propose to focus on only one dimension at a time. At each iteration, we select a 

% Instead of searching for the acquisition function optimizer all over the design space, we select one-dimensional subspace at a time to 

\subsection{Effective Dimension Selection} \label{sec:dimension_selection}

According to the above, another interesting question arises: how to select the effective dimension for optimization? 

\textbf{Random Selection:} The intuitive and naive way of dimension selection is random selection. By randomly selecting the dimension for optimization, the design space at each one-dimensional subspace can be equally searched, thus, globally explored. However, this dimension selection policy greatly neglects the characteristics of different optimization problems and tends to evenly sample around the low and high potential subspaces.

\textbf{Descent Direction Estimation:} Another dimension selection policy is to calculate the descent direction heuristically based on the Thompson sampling \cite{thompson1933likelihood}. By evaluating the utility of the best data point $\bm{x}_\ast$ and the $d$ data points around it with a small deviation $\bm{x}_\ast + \hat{g} l_i$ in each direction, we can calculate the gradient $\hat{g}$ at each dimension and focus on the most crucial one. However, this dimension selection strategy tends to constantly sample around several most effective subspaces and easily get stuck in the local optimum.

In this paper, we propose a mixture of selection strategy to combine the benefits of random selection and descent direction estimation. With probability $p$, we randomly select the dimension for optimization. With probability $1-p$, we select the dimension with descent direction estimation based on the Thompson sampling. In this way, parameter $p$ helps to balance between searching globally or locally. And we fix $p$ as 0.8 in this paper. 

\subsection{Summary} \label{sec:summary}

Instead of optimizing the acquisition function over the whole design space, we propose to focus on a single dimension at a time to ensure information gain at each iteration. By combining the random selection strategy and the descent direction estimation policy, we adaptively avoid sampling too much in the low potential subspaces while globally exploring the design space. To fully utilize the hardware resources and further accelerate the optimization procedure, we also leverage the asynchronous batch Bayesian optimization framework EasyBO \cite{zhang2020bayesian} to enable parallelism. By randomly selecting the exploration parameter, we achieve a delicate exploration-exploitation tradeoff and maximize the information gain per data point. The heuristically observed data points are employed to naturally penalize around the previously selected data points that are still under observation. The overall framework of our proposed algorithm LinEasyBO is presented in Algorithm \ref{algo:LinEasyBO}.

\begin{algorithm}
    \centering
    \caption{LinEasyBO Framework}
    \label{algo:LinEasyBO}
    \begin{algorithmic}
        \STATE Randomly sample an initial dataset $D_0=\{X, \bm{y}\}$, and the number of works $B$ to evaluate the design in parallel
        \FOR{t = 1 to T}
        \STATE Wait for a worker to finish
        % \STATE Observe the selected data $y_t = f(\bm{x}_t)$
        % \STATE Update the training dataset $D_t = \{D_{t-1}, \{\bm{x}_t, y_t\}\}$
        \STATE Query the observed data $D_t = \{D_{t-1}, \{\bm{x}_t, y_t\}\}$
        \STATE Build a Gaussian process regression model with $D_t \cup \{\hat{X}, \hat{\bm{y}}\}$, where $\hat{X} = \{\hat{\bm{x}}_1, \cdots, \hat{\bm{x}}_{B-1}\}$ is the previously selected data points remaining to be observed, and $\hat{\bm{y}} = \{\mu(\hat{\bm{x}}_1), \cdots, \mu(\hat{\bm{x}}_{B-1})\}$ is the predictive mean of the $\hat{X}$ that works as the heuristically observated data points
        \STATE Get the current best design $\bm{x}_\ast = \text{argmax}_{\bm{x}} f(\bm{x})$
        \STATE Select the dimension $l_t$ for optimization (\S\ref{sec:dimension_selection})
        \STATE Select $\bm{x}_t$ that optimizes the acquisition function (\S\ref{sec:linebo})
        \STATE Observe the selected data with the idle worker $y_t = f(\bm{x}_t)$
        \ENDFOR
        \RETURN The best $f(\bm{x})$ recorded during optimization
    \end{algorithmic}
\end{algorithm}

\section{Experimental Results} \label{sec:experiments}

% In this section, we conduct experiments on two real-world analog circuits and compare LinEasyBO with several state-of-the-art optimization algorithms. In general, the tested algorithms can be divided into three categories: the baseline Bayesian optimization algorithms that work in sequential mode, the sequential Bayesian optimization algorithms that selects the query points in one-dimensional subspaces, and the Bayesian optimization algorithms that work in batch mode.

In this section, we conduct experiments on two real-world analog circuits: the charge pump circuit (\S\ref{sec:charge_pump}) and the class-E power amplifier circuit (\S\ref{sec:classE}). To demonstrate the efficiency of our proposed method in both sequential and batch mode, we compared the performances of LinEasyBO with several state-of-the-art algorithms in terms of both optimization results and wall-clock time. To distinguish between the sequential and batch mode, we label each algorithm with its corresponding batch size. In general, the tested algorithms can be divided into the following four categories. 
\begin{enumerate}[label=\textbf{C\arabic*},topsep=2pt,itemsep=1pt,partopsep=2pt, parsep=2pt, leftmargin=*]
\item Random search, a perfect algorithm to measure the difficulty of optimization problems and the efficiency of the optimization algorithms \cite{bergstra2012random}. 
\item Sequential Bayesian optimization algorithms that work as baselines.
\item Sequential Bayesian optimization algorithms that selects the query points in one-dimensional subspaces including our proposed algorithm.
\item Bayesian optimization algorithms that work in the batch mode.
\end{enumerate}

% All experiments are conducted on a Linux workstation with two Intel Xeon CPUs and 128GB memory. All circuit performances are generated with commercial HPSICE circuit simulator. Our proposed algorithm LinEasyBO is compared with several state-of-the-art optimization algorithms in terms of both optimization results and wall-clock time.

% By comparing our proposed method with several state-of-the-art optimization algorithms, we demonstrate the efficiency of LinEasyBO in terms of both optimization results and wall-clock time. 

% For Bayesian optimization algorithms that work in sequential mode, we run experiments on three algorithms:
For the sequential Bayesian optimization algorithms that work as baselines (\textbf{C2}), we run experiments on three algorithms:
(1) EI \cite{mockus1978application}, an improvement-based acquisition function that helps to explore the design space.
(2) LCB \cite{srinivas2009gaussian}, an optimistic policy that searches the design space in sequence.
(3) EasyBO \cite{zhang2020bayesian}, an efficient Bayesian optimization algorithm that selects the exploration parameter randomly.

To demonstrate the effectiveness of optimizing the inner loop in the one-dimensional subspaces (\textbf{C3}), we run experiments on three alternative algorithms:
(1) Line-EI, an improvement-based policy that optimizes the acquisition function in one-dimensional subspaces.
(2) Line-LCB, an optimistic strategy that searches the next query points in line.
(3) LinEasyBO, which is our proposed algorithm in sequential mode.

Finally, we comprehensively instantiate \textbf{C4} with experiments on four algorithms:
(1) LP-EI \cite{gonzalez2016batch}, an improvement-based strategy that maintains diversity in the batch with Lipschitz constant.
(2) LP-LCB \cite{gonzalez2016batch}, an optimistic policy that introduces a local repulsive term to reduce redundantly sampling around the same area.
(3) REMBOpBO \cite{hu2019enabling}, an efficient Bayesian optimization framework that handles the high-dimensional problems by projecting the design variables into a high effective subspace.
(4) LinEasyBO, which is our proposed asynchronous algorithm in batch mode.

% an  batch Bayesian optimization algorithm based on the expected improvement function that maintains diversity within the batch with a local repulsion term.
% Considering that random search is a perfect algorithm to measure the difficulty of optimization problems and the efficiency of the optimization algorithms \cite{bergstra2012random}, we also run experiments on random search algorithm. To distinguish between the sequential and batch mode, we label each algorithm with its corresponding batch size.

To ensure a fair comparison and reduce implementation bias, we implement the Gaussian process regression model with GPyOpt \cite{gonzalez2016batch} library for all test algorithms. For REMBOpBO, we follow the experimental setting of \cite{hu2019enabling} and optimize the acquisition function with NLopt \cite{johnson2014nlopt} library. For the rest of the algorithms, we implement the inner optimization procedure with Scipy \cite{jones2001scipy} library. For each algorithm, we limit the maximum number of circuit simulation as 350 and set the size of the initial dataset as 20. We also run each algorithm 20 times to reduce random fluctuations. All circuit performances are generated with the commercial HSPICE circuit simulator. And all experiments are conducted on a Linux workstation with two Intel Xeon CPUs and 128GB memory.

\subsection{Charge Pump} \label{sec:charge_pump}

\begin{figure}
    \vspace{-3mm}
    \centering
    \includegraphics[width=0.45\textwidth]{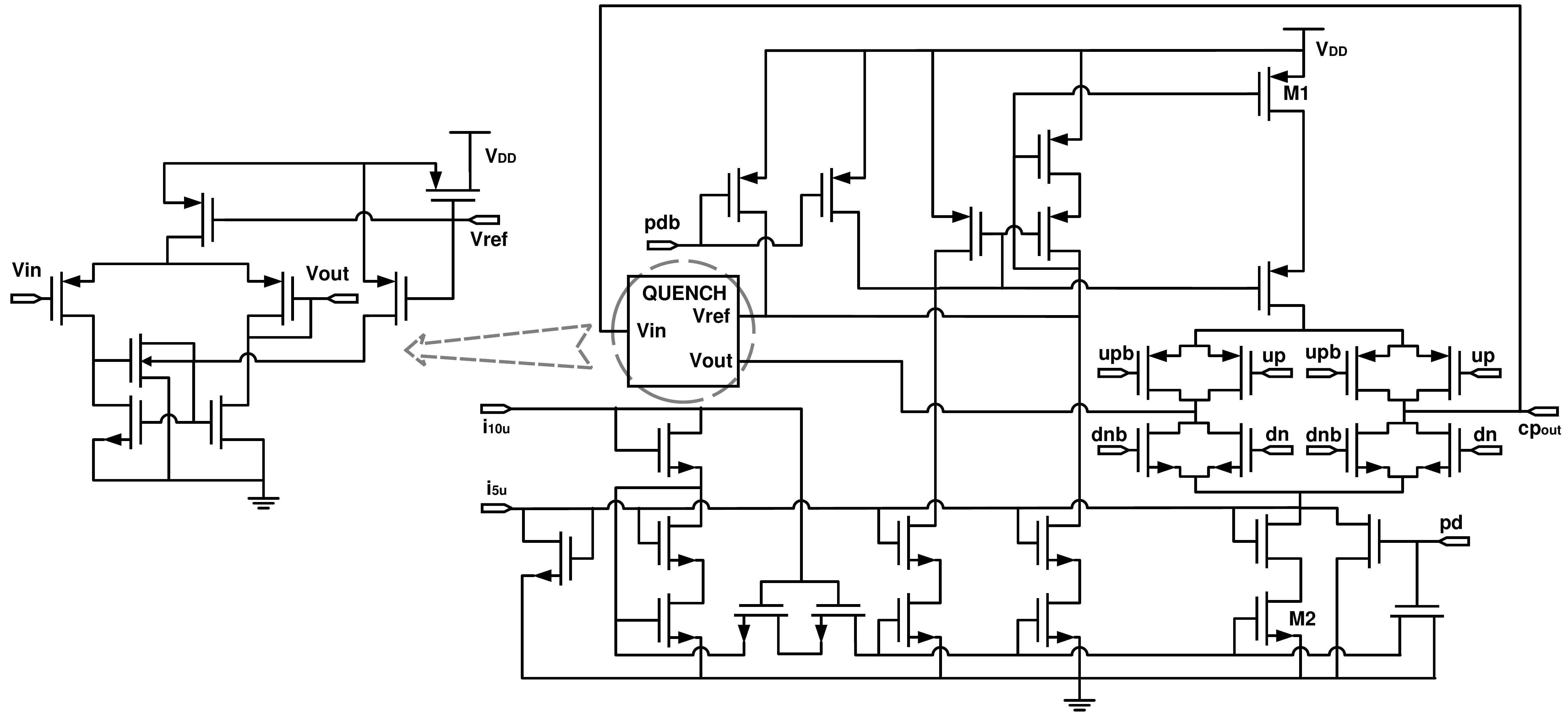}
    \caption{The schematic of the charge pump circuit, which is reproduced from \cite{yang2018smart-msp:}.\vspace{-3mm}}
    \label{fig:charge_pump}
\end{figure}

The schematic of the charge pump circuit is presented in Figure \ref{fig:charge_pump}. Implemented in an SMIC 40nm process, there are a total of 36 design variables, including lengths and widths of the transistors, the resistance of the resistors and the capacitance of the capacitors. The target of design is to ensure the current of M1 and M2 stay close to 40$\mu A$ across 18 PVT corners \cite{yang2018smart-msp:}. The design specification of the charge pump is as follows:
\begin{equation}
    \label{eq:charge_pump1}
    \text{minimize} \quad 0.3 \times \mathit{diff} + 0.5 \times \mathit{deviation}, 
\end{equation}
where 
\begin{equation}
    \label{eq:charge_pump2}
    \begin{cases}
        \begin{aligned}
            \mathit{diff_1} & =  \text{max}_{\forall PVT} (I_{M1, max} - I_{M1, avg}), \\
            \mathit{diff_2} & =  \text{max}_{\forall PVT} (I_{M1, avg} - I_{M1, min}), \\
            \mathit{diff_3} & = \text{max}_{\forall PVT} (I_{M2, max} - I_{M2, avg}), \\
            \mathit{diff_4} & =  \text{max}_{\forall PVT} (I_{M2, avg} - I_{M2, min}), \\
            \mathit{diff} & =  \sum^{4}_{i=1} \mathit{diff_{i}}, \\
            \mathit{deviation} & =  \text{max}_{\forall PVT} (I_{M1, avg} - 40\mu A) \\
                   & +  \text{max}_{\forall PVT} (I_{M2, avg} - 40\mu A). \\
        \end{aligned}
    \end{cases}
\end{equation}
And the experimental results are presented in Table \ref{tb:charge_pump}.

\begin{table}
    \centering
    \caption{The optimization results of the charge pump circuit.}
    \label{tb:charge_pump}
    \begin{tabular}{cccccc}
        \hline
        Algo & Best & Worst & Mean & Std & Time \\
        \hline
        Random Search & 21.16 & 46.39 & 30.49 & 7.52 & - \\
        EI & 3.42 & 4.67 & 4.09 & 0.35 & 7h45m32s\\
        LCB & 3.45 & 27.92 & 5.22 & 5.22 & 6h34m11s\\
        EasyBO & 5.09 & 7.43 & 6.15 & 0.69 & 7h57m51s\\
        \hline
        Line-EI & 3.11 & 4.40 & 3.37 & 0.29 & 7h56m52s\\
        Line-LCB & 3.16 & 3.77 & 3.39 & 0.17 & 7h53m1s\\
        LinEasyBO & 3.16 & 4.06 & 3.39 & 0.22 & 7h39m49s\\
        \hline
        LP-EI-5 & 3.49 & 5.13 & 4.06 & 0.39 & 2h10m54s\\
        LP-LCB-5 & 3.63 & 5.08 & 4.17 & 0.36 & 1h48m6s\\
        REMBOpBO-5 & 9.86 & 32.75 & 14.70 & 4.87 & 2h21m48s \\
        LinEasyBO-5 & 3.18 & 4.11 & \textbf{3.44} & 0.25 & \textbf{1h45m58s}\\
        \hline
        LP-EI-10 & 3.77 & 5.63 & 4.59 & 0.45 & 1h33m38s\\
        LP-LCB-10 & 3.53 & 4.89 & 4.24 & 0.30 & 53m52s\\
        REMBOpBO-10 & 8.76 & 27.68 & 17.51 & 5.12 & 1h43m2s\\
        LinEasyBO-10 & 3.32 & 4.71 & \textbf{3.71} & 0.34 & \textbf{49m58s}\\
        \hline
        LP-EI-15 & 4.02 & 5.85 & 5.02 & 0.49 & 1h22m23s\\
        LP-LCB-15 & 3.60 & 4.93 & 4.29 & 0.33 & 42m31s\\
        REMBOpBO-15 & 11.35 & 32.12 & 18.31 & 5.62 & 1h28m2s\\
        LinEasyBO-15 & 3.23 & 4.90 & \textbf{3.95} & 0.41 & \textbf{33m33s}\\
        \hline
        % randomly select beta parameter from [0, 14]
        % LinEasyBO14-5 & 3.15 & 4.12 & 3.51 & 0.28 & \\
        % LinEasyBO14-10 & 3.34 & 3.91 & 3.61 & 0.16 & \\
        % LinEasyBO14-15 & 3.40 & 4.30 & 3.83 & 0.24 & \\
    \end{tabular}
\end{table}

Let us first focus on the sequential mode. It is worth notice that the performances of EI, LCB and EasyBO significantly outperform that of the random search algorithm. This reveals that the design specification of the charge pump circuit is hard to optimization and the Bayesian optimization framework can guide the search efficiently and effectively. Another interesting observation is that those optimization algorithms that only focus on exploring one dimension at a time (Line-EI, Line-LCB and LinEasyBO) consistently outperform their traditional counterparts. This clearly suggests that searching the design space in one-dimensional subspaces is beneficial regardless of the acquisition function. This also emphasizes that the efficiency of the Bayesian optimization framework can be largely impacted by the inner optimization procedure. By solely focusing on one dimension at a time, space exploration efficiency can be greatly improved.

% the efficiency of the inner optimization procedure can greatly

% By solely focusing on direction at a time, the space exploration efficiency can be safely guaranteed 

% the performances of EI, LCB and EasyBO can be greatly improved by focusing on exploring one dimension at a time.

% Let us first focus on the sequential mode. It is worth notice that the performances of EI, LCB and EasyBO are much better than that of the random search algorithm. This reveals that the design specification of the charge pump circuit is hard to optimization. Also, the Bayesian optimization approaches (including EI, LCB, EasyBO) consistently outperforms random search also shows that the Bayesian optimization algorithm can better guide the search and greatly increase the sampling efficiency. Another interesting observation is that optimization algorithms that only focus on one-dimensional subspaces at each iteration significantly outperform their traditional counterparts that search the global optimum of the acquisition function on the whole. This shows that the information gain per observed data point can be greatly improved by simply increasing the effectiveness of the acquisition function optimizer. By solely focusing on one-dimensional subspaces, the space exploration efficiency can be greatly improved in terms of both sampling efficiency and wall-clock time.

\begin{figure}
    \centering
    \vspace{-5mm}
    \includegraphics[width=0.45\textwidth]{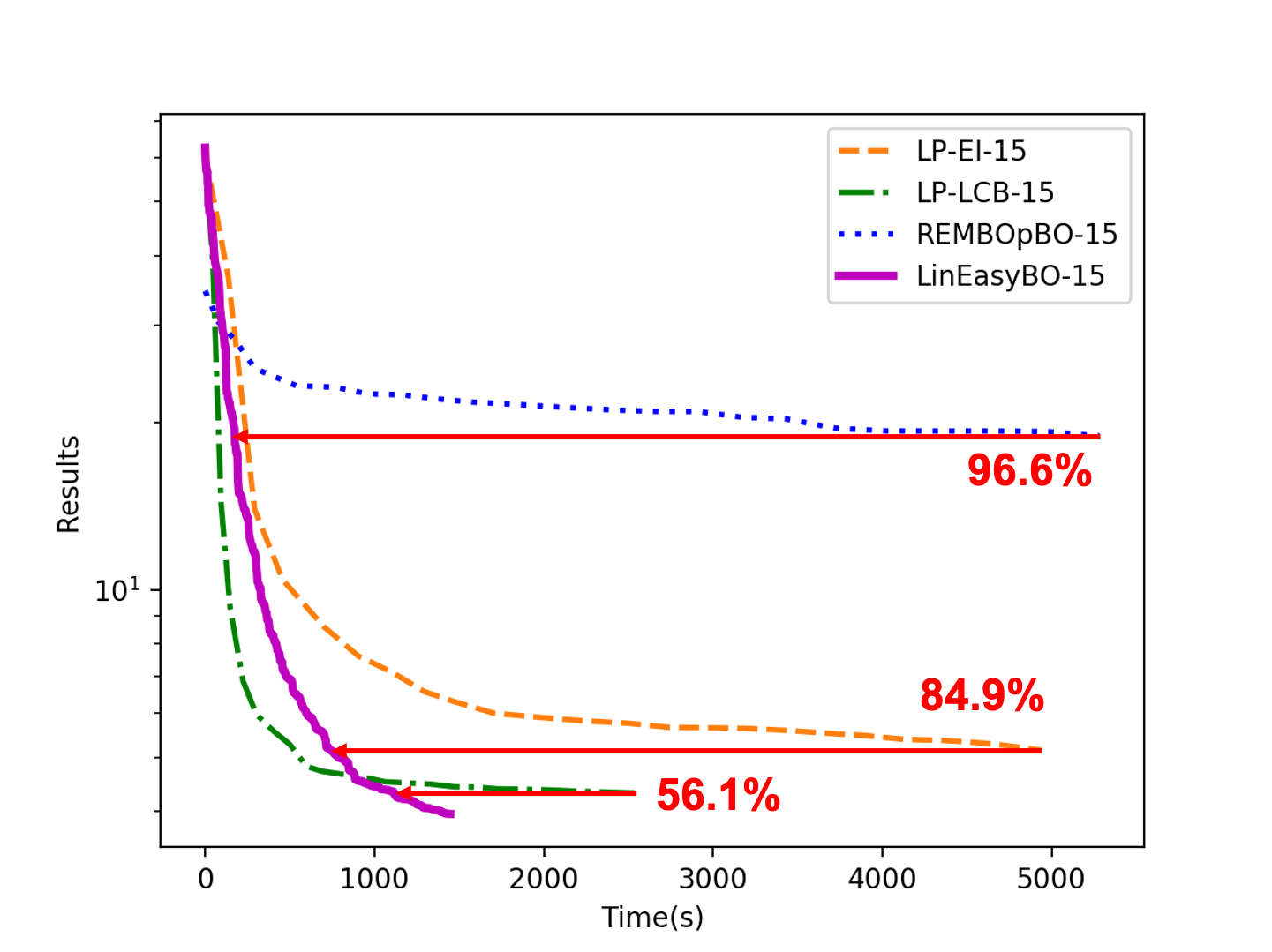}
    \caption{The optimization results with respect to the wall clock time for the charge pump circuit, when the batch size is 15.\vspace{-3mm}}
    \label{fig:charge_pump_results}
\end{figure}

Now let us move on to analyzing the optimization results in batch mode. Compared to LP-EI, LP-LCB, and REMBOpBO, LinEasyBO consistently demonstrates better performances in terms of the optimization results and time consumption on average regardless of the batch size. This clearly shows the efficiency and effectiveness of our proposed algorithm. The performances of LP-EI, LP-LCB, REMBOpBO, and LinEasyBO deteriorate with the increase of the batch size. This implies that the design specification of the charge pump circuit is rather sensitive to the exploration parameter of the acquisition function. In other words, even if there is only a small change in the exploration-exploitation tradeoff of the acquisition function, the final optimization results can fluctuate greatly for this circuit. The fact that LinEasyBO constantly outperforms EI, LCB and EasyBO in both sequential and batch mode again emphasizes the importance of the inner optimization procedure. The fact that REMBOpBO is always outperformed to a large extend by LP-EI, LP-LCB, and LinEasyBO suggests that the low effective dimensionality preassumption doesn't hold for the charge pump circuit. The strong assumption that a low-dimensional effective subspace exists for the optimization problem can greatly limit the application of REMBOpBO, while LinEasyBO is not affected by the characteristic of the problem and can be widely and safely used without much consideration of the application scenario. As is shown in Figure \ref{fig:charge_pump_results}, compared with LP-LCB, LP-EI, and REMBOpBO, LinEasyBO reduces 56.1\%, 84.9\%, and 96.6\% of the time consumption respectively when the batch size is 15, while achieving the same optimization results. This translates to $2\times$, $6\times$, and $29\times$ optimization procedure acceleration respectively.

% LinEasyBO consistently obtains better optimization results and less time consumption
% It is obvious that LinEasyBO consistently demonstrates better optimization results and less time consumption regardless of the batch size 

% Our proposed algorithm LinEasyBO constantly outperforms the state-of-the-art batch algorithms regardless of the batch size. This further demonstrates the efficiency and effectiveness of our proposed algorithm. The performance of LP-EI, LP-LCB, REMBOpBO and LinEasyBO deteriorate with the increase of the batch size. This implies that the design specification of the charge pump circuit is rather sensitive to the exploration parameter. In other words, even if there is only a small change in the exploration-exploitation tradeoff of the acquisition function, the final optimization results of the optimization algorithms can fluctuate greatly for this circuit. However, the performance of LinEasyBO outperforms EI, LCB and EasyBO regardless of the batch and sequential. This emphasises the importance of the acquisition function optimization procedure. The fact that REMBOpBO is constantly outperformed by LinEasyBO shows that all design variables in the charge pump circuit counts and the preassumption that only several design variables have a great influence on the circuit performance doesn't hold for every circuit. Also, this strong preassumption will greatly limit the application of REMBOpBO while the performance of LinEasyBO is not affected by the characteristic of the problem and can be widely used without much consideration of the application scenario.

\subsection{Class-E Power Amplifier} \label{sec:classE}

\begin{figure}
    \centering
    \includegraphics[width=0.45\textwidth]{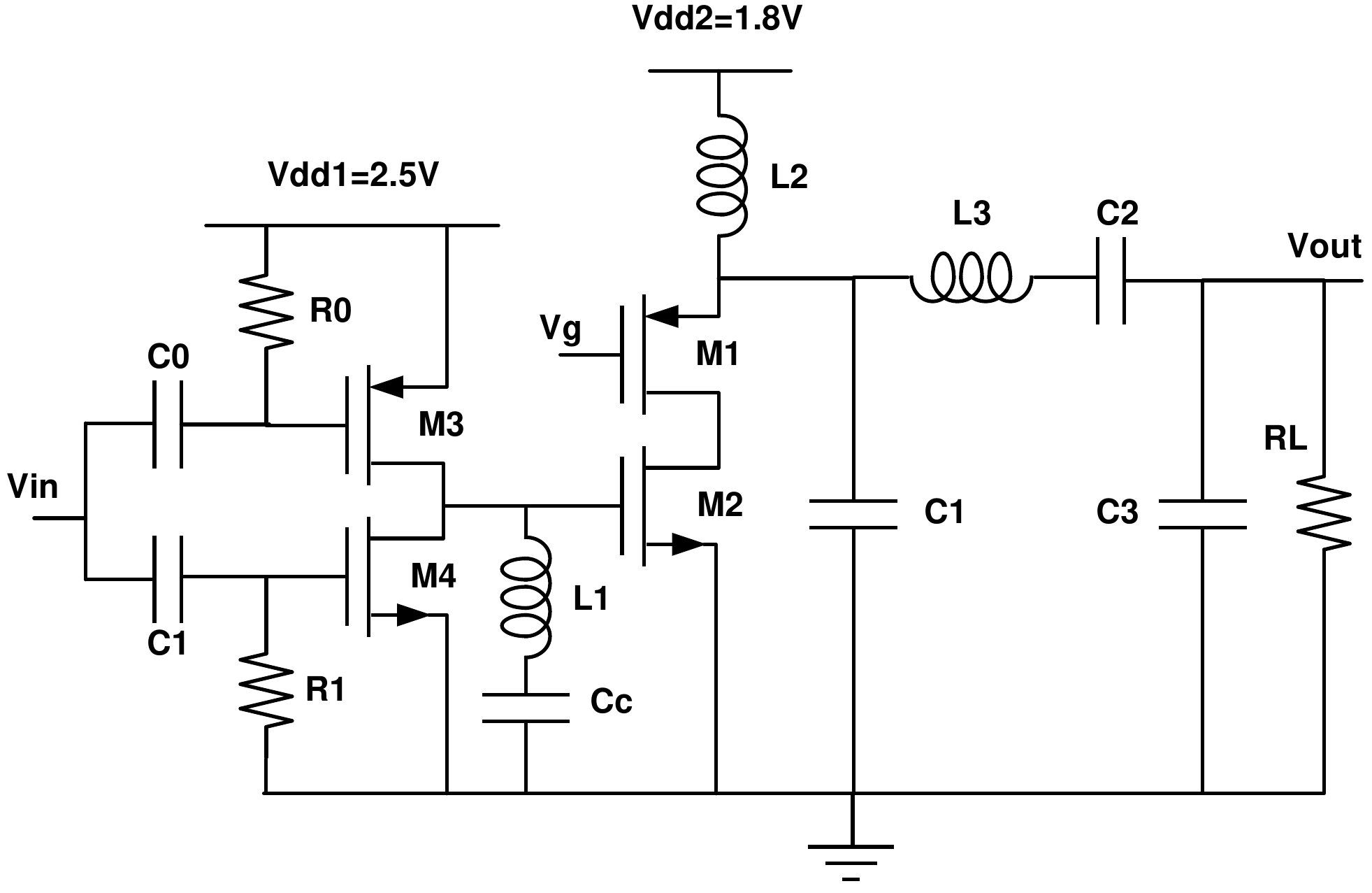}
    \caption{The schematic of the class-E power amplifier circuit, which is reproduced from \cite{lyu2018batch}.\vspace{-2mm}}
    \label{fig:classE}
\end{figure}

The schematic of the class-E power amplifier circuit is presented in Figure \ref{fig:classE}. Implemented in a TSMC 180nm process, there are a total of 12 design variables. The target of design for this circuit is to maximize the power-added efficiency (PAE) and the output power (Pout) simultaneously \cite{lyu2018batch}. The corresponding design specification is as follows:
\begin{equation}
    \label{eq:classE}
    \text{maximize} \quad 3 \times PAE + Pout.
\end{equation}
And the optimization results are shown in Table \ref{tb:classE}.

As usual, we start with looking at the sequential mode. Again, the Bayesian optimization algorithms with our proposed acquisition function optimizer consistently outperform their traditional counterparts. This again demonstrates the efficiency and effectiveness of searching the design space one dimension at a time and the importance of successfully optimizing the acquisition function at each iteration. Another interesting and unexpected phenomenon is that random search algorithm obtains better optimization results than EI and comparable performance with LCB. This strongly stresses the fact that a less efficient inner optimization strategy can guide the search in the wrong direction and greatly hurt the performance of the optimization algorithm. By only focusing on a single dimension and optimizing the acquisition function optimally at each iteration, the Bayesian optimization algorithm can iteratively make solid progress and search the state space in a safe and efficient manner. Our proposed acquisition function optimization method can be easy-to-use and effective regardless of the acquisition function and beneficial across different dimensionality.

% Let us first focus on the sequential mode. Again the Bayesian optimization algorithms with refined acquisition function optimizer consistently outperform their traditional counterparts. This again suggests the effectiveness of the RandomLineBO strategy and the importance of successfully finding the global optimum of the acquisition function at each iteration. Another interesting phenomenon is that the random search algorithm has better performance than EI and comparable performance with LCB. This strongly demonstrates that the efficiency of the Bayesian optimization approaches can be greatly hurt if the acquisition function optimizer cannot guarantee to reach the global optimum at each iteration. Therefore, by only focusing on a single dimension and searching the global optimum within that dimensional subspace at each iteration, the Bayesian optimization algorithm can iteratively make solid progress at each iteration, thus, can search the state space in a much safer and efficient manner. With the increase of the dimensionality of the problem, the performance of the Bayesian optimization algorithm will be less hurt by the curse of dimensionality. By focusing on only one-dimensional subspace and searching for the global optimum of the acquisition function, the optimization efficiency can be increased regardless of the acquisition functions.

\begin{figure}
    \centering
    \vspace{-3mm}
    \includegraphics[width=0.45\textwidth]{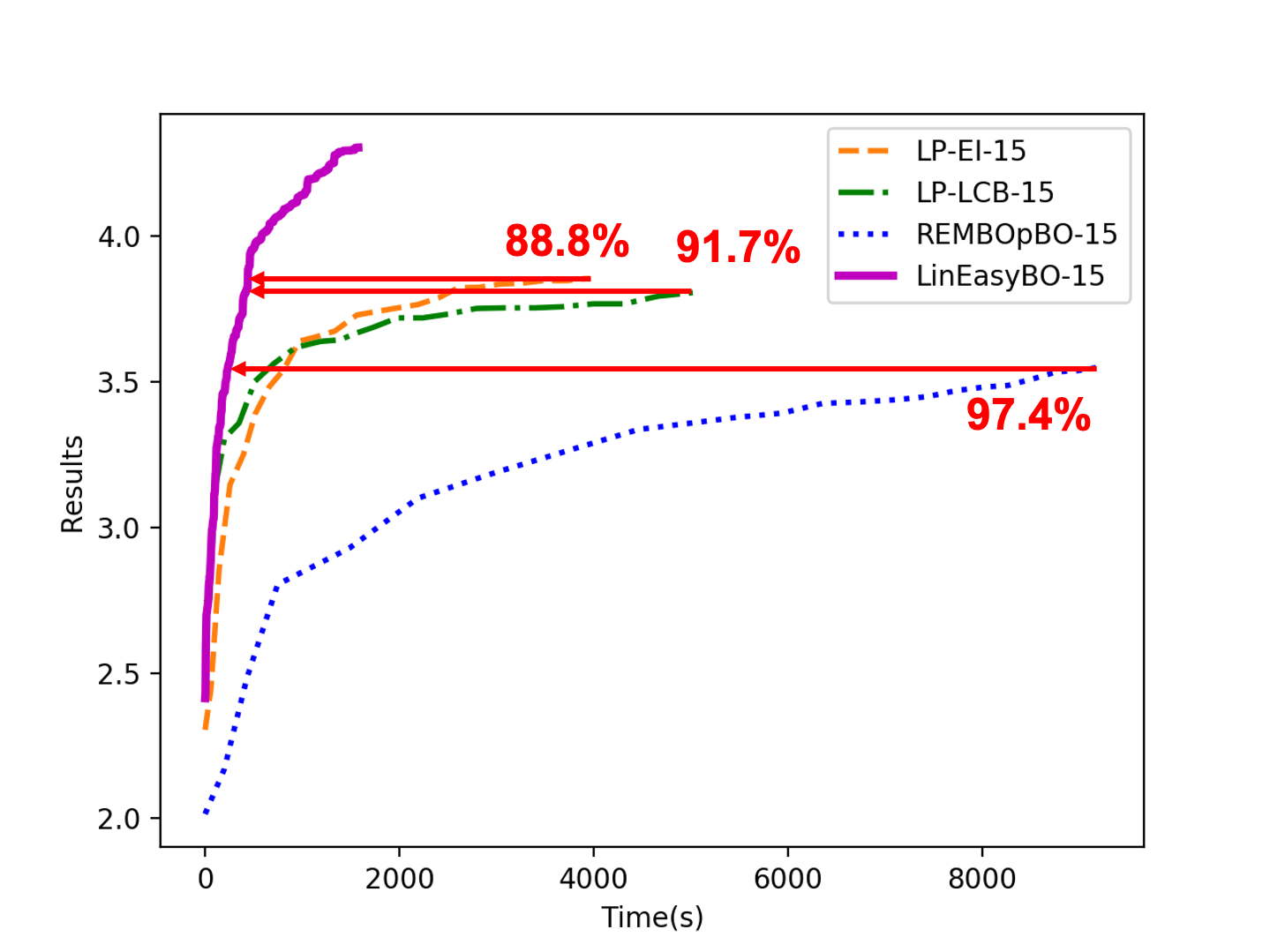}
    \caption{The optimization results on average with respect to the wall-clock time for the class-E power amplifier circuit, when the batch size is 15.\vspace{-2mm}}
    \label{fig:classE_results}
\end{figure}

Let's now analyze the optimization results in batch mode. One interesting phenomenon is that REMBOpBO is constantly outperformed by random search. The assumption that only a small group of design variables have a great impact on the circuit performance can significantly hurt the performance of REMBOpBO when the preassumption is not met. This makes REMBOpBO a risky choice when we have no information at hand about the black-box optimization problem. The fact that LinEasyBO consistently outperforms LP-EI, LP-LCB, and REMBOpBO with less time consumption again demonstrates the efficiency and effectiveness of our proposed algorithm. By focusing on a single dimension at a time, the acquisition function optimization procedure can help to make solid and safe decisions and maximize the information gain in the long run. The comparable optimization results of LinEasyBO across different batch size show the robustness of our proposed algorithm. As is presented in Figure \ref{fig:classE_results}, LinEasyBO reduces the simulation time (accelerates the optimization procedure) by 88.8\% ($9\times$), 91.7\% ($11\times$), and 97.4\% ($38\times$) compared to LP-EI, LP-LCB, and REMBOpBO respectively when the batch size is 15, while obtaining the same optimization results.

% only a small group of design variables have great impact on the circuit performance is a strong preassumption and 

% Now we move on analyzing the batch mode. One interesting phenomenon is that the performance of REMBOpBO are constantly outperformed by the random search algorithm. This again suggests that only a small group of design variables have great impact on the circuit performance is a strong preassumption and this preassumption will greatly limit the application of REMBOpBO. Also, the performance of REMBOpBO will be greatly hurt if this preassumption doesn't hold. The fact that LinEasyBO outperforms all the rest of the algorithms demonstrates the efficiency and effectiveness of our proposed algorithm. By making solid and safe progress at each iteration, the optimization efficiency can be greatly increased in the long run. LinEasyBO also achieves comparable optimization results across different batch size. This demonstrates the robustness of our proposed algorithm.

\begin{table}
    \centering
    \caption{The optimization results of class-E power amplifier circuit.}
    \label{tb:classE}
    \begin{tabular}{cccccc}
        \hline
        Algo & Best & Worst & Mean & Std & Time \\
        \hline
        Random Search & 4.16 & 3.34 & 3.70 & 0.21 & - \\
        EI & 4.36 & 1.60 & 3.68 & 0.52 & 12h22m8s\\
        LCB & 4.12 & 3.42 & 3.79 & 0.21 & 9h10m31s\\
        EasyBO & 4.51 & 3.69 & 4.01 & 0.15 & 6h49m46s\\
        \hline
        Line-EI & 4.83 & 3.21 & 4.29 & 0.49 & 6h4m45s \\
        Line-LCB & 4.81 & 2.82 & 4.03 & 0.52 & 6h4m46s\\
        LinEasyBO & 5.44 & 3.35 & 4.34 & 0.58 & 6h17m15s\\
        \hline
        LP-EI-5 & 4.17 & 3.35 & 3.83 & 0.21 & 1h45m3s\\
        LP-LCB-5 & 4.26 & 3.42 & 3.79 & 0.26 & 2h1m38s\\
        REMBOpBO-5 & 4.08 & 3.04 & 3.57 & 0.21 & 2h39m12s \\
        LinEasyBO-5 & 5.53 & 3.19 & \textbf{4.32} & 0.47 & \textbf{1h21m41s}\\
        \hline
        LP-EI-10 & 4.60 & 3.39 & 3.84 & 0.30 & 1h17m14s\\
        LP-LCB-10 & 4.21 & 3.45 & 3.77 & 0.20 & 1h34m13s\\
        REMBOpBO-10 & 3.92 & 1.73 & 3.54 & 0.46 & 2h44m57s\\
        LinEasyBO-10 & 5.32 & 3.16 & \textbf{4.25} & 0.57 & \textbf{40m43s} \\
        \hline
        LP-EI-15 & 4.34 & 3.49 & 3.86 & 0.23 & 1h6m3s\\
        LP-LCB-15 & 4.08 & 3.41 & 3.81 & 0.15 & 1h23m40s\\
        REMBOpBO-15 & 4.10 & 2.92 & 3.55 & 0.22 & 2h33m23s\\
        LinEasyBO-15 & 5.96 & 3.21 & \textbf{4.30} & 0.74 & \textbf{27m23s} \\
        \hline
    \end{tabular}
\end{table}

\section{Conclusion} \label{sec:conclusion}

In this paper, we proposed a fast and robust Bayesian optimization framework for analog circuit synthesis via one-dimensional subspaces. Instead of searching the global optimum for acquisition function over the whole design space, our proposed method focuses on one dimension at a time. We also introduce an effective dimension selection strategy to balance searching globally and locally. In this way, our proposed algorithm guarantees information gain at each step and searches the design space safely and efficiently. 
To further accelerate the optimization procedure, we leverage the batch Bayesian optimization framework to fully utilize the computational resources. Compared to the traditional acquisition function optimization procedure, our method not only reduces the computational overhead for the Bayesian optimization framework but also increases the information gain per observation. We experimentally compare the performance of our proposed algorithm with several state-of-the-art optimization algorithms. Experimental results quantitatively show that our method accelerates the optimization procedure by up to $9\times$ and $38\times$ compared to LP-EI and REMBOpBO when the batch size is 15.

\vspace{-1mm}
% \section*{Acknowledgment}
% \section*{References}
\bibliographystyle{IEEEtran}
\bibliography{include/bibliography}

\begin{thebibliography}{10}
\providecommand{\url}[1]{#1}
\csname url@samestyle\endcsname
\providecommand{\newblock}{\relax}
\providecommand{\bibinfo}[2]{#2}
\providecommand{\BIBentrySTDinterwordspacing}{\spaceskip=0pt\relax}
\providecommand{\BIBentryALTinterwordstretchfactor}{4}
\providecommand{\BIBentryALTinterwordspacing}{\spaceskip=\fontdimen2\font plus
\BIBentryALTinterwordstretchfactor\fontdimen3\font minus
  \fontdimen4\font\relax}
\providecommand{\BIBforeignlanguage}[2]{{%
\expandafter\ifx\csname l@#1\endcsname\relax
\typeout{** WARNING: IEEEtran.bst: No hyphenation pattern has been}%
\typeout{** loaded for the language `#1'. Using the pattern for}%
\typeout{** the default language instead.}%
\else
\language=\csname l@#1\endcsname
\fi
#2}}
\providecommand{\BIBdecl}{\relax}
\BIBdecl

\bibitem{rutenbar2007hierarchical}
R.~A. Rutenbar, G.~G. Gielen, and J.~Roychowdhury, ``Hierarchical modeling,
  optimization, and synthesis for system-level analog and rf designs,''
  \emph{Proceedings of the IEEE}, vol.~95, no.~3, pp. 640--669, 2007.

\bibitem{liu2009analog}
B.~Liu, Y.~Wang, Z.~Yu, L.~Liu, M.~Li, Z.~Wang, J.~Lu, and F.~V. Fernandez,
  ``Analog circuit optimization system based on hybrid evolutionary
  algorithms,'' \emph{Integration}, vol.~42, no.~2, pp. 137--148, 2009.

\bibitem{liu2014gaspad:}
B.~Liu, D.~Zhao, P.~Reynaert, and G.~Gielen, ``Gaspad: A general and efficient
  mm-wave integrated circuit synthesis method based on surrogate model assisted
  evolutionary algorithm,'' \emph{IEEE Transactions on Computer-Aided Design of
  Integrated Circuits and Systems}, vol.~33, no.~2, pp. 169--182, 2014.

\bibitem{phelps2000anaconda:}
R.~Phelps, M.~Krasnicki, R.~A. Rutenbar, L.~R. Carley, and J.~R. Hellums,
  ``Anaconda: simulation-based synthesis of analog circuits via stochastic
  pattern search,'' \emph{IEEE Transactions on Computer-Aided Design of
  Integrated Circuits and Systems}, vol.~19, no.~6, pp. 703--717, 2000.

\bibitem{hu2018parallelizable}
H.~Hu, P.~Li, and J.~Z. Huang, ``Parallelizable bayesian optimization for
  analog and mixed-signal rare failure detection with high coverage,'' in
  \emph{Proceedings of the International Conference on Computer-Aided Design},
  2018, pp. 1--8.

\bibitem{yang2018smart-msp:}
Y.~Yang, H.~Zhu, Z.~Bi, C.~Yan, D.~Zhou, Y.~Su, and X.~Zeng, ``Smart-msp: A
  self-adaptive multiple starting point optimization approach for analog
  circuit synthesis,'' \emph{IEEE Transactions on Computer-Aided Design of
  Integrated Circuits and Systems}, vol.~37, no.~3, pp. 531--544, 2018.

\bibitem{lyu2017efficient}
W.~Lyu, P.~Xue, F.~Yang, C.~Yan, Z.~Hong, X.~Zeng, and D.~Zhou, ``An efficient
  bayesian optimization approach for automated optimization of analog
  circuits,'' \emph{IEEE Transactions on Circuits and Systems I: Regular
  Papers}, vol.~65, no.~6, pp. 1954--1967, 2017.

\bibitem{boyd2005geometric}
S.~P. Boyd and S.~J. Kim, ``Geometric programming for circuit optimization,''
  in \emph{Proceedings of the 2005 international symposium on Physical design},
  2005, pp. 44--46.

\bibitem{boyd2007tutorial}
S.~Boyd, S.-J. Kim, L.~Vandenberghe, and A.~Hassibi, ``A tutorial on geometric
  programming,'' \emph{Optimization and engineering}, vol.~8, no.~1, p.~67,
  2007.

\bibitem{hannah2012ensemble}
L.~Hannah and D.~Dunson, ``Ensemble methods for convex regression with
  applications to geometric programming based circuit design,'' \emph{arXiv
  preprint arXiv:1206.4645}, 2012.

\bibitem{gielen1990analog}
G.~Gielen, H.~Walscharts, and W.~Sansen, ``Analog circuit design optimization
  based on symbolic simulation and simulated annealing,'' \emph{IEEE Journal of
  Solid-state Circuits}, vol.~25, no.~3, pp. 707--713, 1990.

\bibitem{grzechca2007use}
D.~Grzechca, T.~Golonek, and J.~Rutkowski, ``The use of simulated annealing
  with fuzzy objective function to optimal frequency selection for analog
  circuit diagnosis,'' in \emph{2007 14th IEEE International Conference on
  Electronics, Circuits and Systems}.\hskip 1em plus 0.5em minus 0.4em\relax
  IEEE, 2007, pp. 899--902.

\bibitem{peng2016efficient}
B.~Peng, F.~Yang, C.~Yan, X.~Zeng, and D.~Zhou, ``Efficient multiple starting
  point optimization for automated analog circuit optimization via recycling
  simulation data,'' in \emph{Proceedings of the 2016 Conference on Design,
  Automation \& Test in Europe}.\hskip 1em plus 0.5em minus 0.4em\relax EDA
  Consortium, 2016, pp. 1417--1422.

\bibitem{alpaydin2003evolutionary}
G.~Alpaydin, S.~Balkir, and G.~Dundar, ``An evolutionary approach to automatic
  synthesis of high-performance analog integrated circuits,'' \emph{IEEE
  Transactions on Evolutionary Computation}, vol.~7, no.~3, pp. 240--252, 2003.

\bibitem{fakhfakh2010analog}
M.~Fakhfakh, Y.~Cooren, A.~Sallem, M.~Loulou, and P.~Siarry, ``Analog circuit
  design optimization through the particle swarm optimization technique,''
  \emph{Analog Integrated Circuits and Signal Processing}, vol.~63, no.~1, pp.
  71--82, 2010.

\bibitem{wu2009a}
C.~Wu, D.~Wang, A.~W.~H. Ip, D.~Wang, C.~Chan, and H.~Wang, ``A particle swarm
  optimization approach for components placement inspection on printed circuit
  boards,'' \emph{Journal of Intelligent Manufacturing}, vol.~20, no.~5, pp.
  535--549, 2009.

\bibitem{shahriari2015taking}
B.~Shahriari, K.~Swersky, Z.~Wang, R.~P. Adams, and N.~De~Freitas, ``Taking the
  human out of the loop: A review of bayesian optimization,'' \emph{Proceedings
  of the IEEE}, vol. 104, no.~1, pp. 148--175, 2015.

\bibitem{brochu2010tutorial}
E.~Brochu, V.~M. Cora, and N.~De~Freitas, ``A tutorial on bayesian optimization
  of expensive cost functions, with application to active user modeling and
  hierarchical reinforcement learning,'' \emph{arXiv preprint arXiv:1012.2599},
  2010.

\bibitem{srinivas2009gaussian}
N.~Srinivas, A.~Krause, S.~M. Kakade, and M.~Seeger, ``Gaussian process
  optimization in the bandit setting: No regret and experimental design,''
  \emph{arXiv preprint arXiv:0912.3995}, 2009.

\bibitem{wang2017max}
Z.~Wang and S.~Jegelka, ``Max-value entropy search for efficient bayesian
  optimization,'' \emph{arXiv preprint arXiv:1703.01968}, 2017.

\bibitem{liu2016multi}
B.~Liu, S.~Koziel, and Q.~Zhang, ``A multi-fidelity surrogate-model-assisted
  evolutionary algorithm for computationally expensive optimization problems,''
  \emph{Journal of computational science}, vol.~12, pp. 28--37, 2016.

\bibitem{zhang2019bayesian}
S.~Zhang, W.~Lyu, F.~Yang, C.~Yan, D.~Zhou, and X.~Zeng, ``Bayesian
  optimization approach for analog circuit synthesis using neural network,'' in
  \emph{2019 Design, Automation \& Test in Europe Conference \& Exhibition
  (DATE)}.\hskip 1em plus 0.5em minus 0.4em\relax IEEE, 2019, pp. 1463--1468.

\bibitem{zhang2019efficient}
S.~Zhang, W.~Lyu, F.~Yang, C.~Yan, D.~Zhou, X.~Zeng, and X.~Hu, ``An efficient
  multi-fidelity bayesian optimization approach for analog circuit synthesis,''
  in \emph{Proceedings of the 56th Annual Design Automation Conference
  2019}.\hskip 1em plus 0.5em minus 0.4em\relax ACM, 2019, pp. 1--6.

\bibitem{zhang2020efficient}
S.~Zhang, F.~Yang, D.~Zhou, and X.~Zeng, ``An efficient asynchronous batch
  bayesian optimization approach for analog circuit synthesis,'' in \emph{2020
  57th ACM/IEEE Design Automation Conference (DAC)}.\hskip 1em plus 0.5em minus
  0.4em\relax IEEE, 2020, pp. 1--6.

\bibitem{lyu2018batch}
W.~Lyu, F.~Yang, C.~Yan, D.~Zhou, and X.~Zeng, ``Batch bayesian optimization
  via multi-objective acquisition ensemble for automated analog circuit
  design,'' in \emph{International Conference on Machine Learning}, 2018, pp.
  3312--3320.

\bibitem{he2020efficient}
B.~He, S.~Zhang, F.~Yang, C.~Yan, D.~Zhou, and X.~Zeng, ``An efficient bayesian
  optimization approach for analog circuit synthesis via sparse gaussian
  process modeling,'' in \emph{2020 Design, Automation \& Test in Europe
  Conference \& Exhibition (DATE)}.\hskip 1em plus 0.5em minus 0.4em\relax
  IEEE, 2020, pp. 67--72.

\bibitem{kirschner2019adaptive}
J.~Kirschner, M.~Mutn{\`y}, N.~Hiller, R.~Ischebeck, and A.~Krause, ``Adaptive
  and safe bayesian optimization in high dimensions via one-dimensional
  subspaces,'' \emph{arXiv preprint arXiv:1902.03229}, 2019.

\bibitem{gonzalez2016batch}
J.~Gonz{\'a}lez, Z.~Dai, P.~Hennig, and N.~Lawrence, ``Batch bayesian
  optimization via local penalization,'' in \emph{Artificial intelligence and
  statistics}, 2016, pp. 648--657.

\bibitem{hu2019enabling}
H.~Hu, P.~Li, and J.~Z. Huang, ``Enabling high-dimensional bayesian
  optimization for efficient failure detection of analog and mixed-signal
  circuits,'' in \emph{2019 56th ACM/IEEE Design Automation Conference
  (DAC)}.\hskip 1em plus 0.5em minus 0.4em\relax IEEE, 2019, pp. 1--6.

\bibitem{rasmussen2003gaussian}
C.~E. Rasmussen, ``Gaussian processes in machine learning,'' \emph{Lecture
  Notes in Computer Science}, pp. 63--71, 2003.

\bibitem{kushner1964new}
H.~J. Kushner, ``A new method of locating the maximum point of an arbitrary
  multipeak curve in the presence of noise,'' 1964.

\bibitem{mockus1978application}
J.~Mockus, V.~Tiesis, and A.~Zilinskas, ``The application of bayesian methods
  for seeking the extremum,'' \emph{Towards global optimization}, vol.~2, no.
  117-129, p.~2, 1978.

\bibitem{thompson1933likelihood}
W.~R. Thompson, ``On the likelihood that one unknown probability exceeds
  another in view of the evidence of two samples,'' \emph{Biometrika}, vol.~25,
  no. 3/4, pp. 285--294, 1933.

\bibitem{hennig2012entropy}
P.~Hennig and C.~J. Schuler, ``Entropy search for information-efficient global
  optimization,'' \emph{The Journal of Machine Learning Research}, vol.~13,
  no.~1, pp. 1809--1837, 2012.

\bibitem{hoffman2011portfolio}
M.~D. Hoffman, E.~Brochu, and N.~de~Freitas, ``Portfolio allocation for
  bayesian optimization.'' in \emph{UAI}.\hskip 1em plus 0.5em minus
  0.4em\relax Citeseer, 2011, pp. 327--336.

\bibitem{wilson2018maximizing}
J.~Wilson, F.~Hutter, and M.~Deisenroth, ``Maximizing acquisition functions for
  bayesian optimization,'' in \emph{Advances in Neural Information Processing
  Systems}, 2018, pp. 9884--9895.

\bibitem{lyu2018multi}
W.~Lyu, F.~Yang, C.~Yan, D.~Zhou, and X.~Zeng, ``Multi-objective bayesian
  optimization for analog/rf circuit synthesis,'' in \emph{Proceedings of the
  55th Annual Design Automation Conference}.\hskip 1em plus 0.5em minus
  0.4em\relax ACM, 2018, pp. 1--6.

\bibitem{jones1993lipschitzian}
D.~R. Jones, C.~D. Perttunen, and B.~E. Stuckman, ``Lipschitzian optimization
  without the lipschitz constant,'' \emph{Journal of optimization Theory and
  Applications}, vol.~79, no.~1, pp. 157--181, 1993.

\bibitem{scarlett2018tight}
J.~Scarlett, ``Tight regret bounds for bayesian optimization in one
  dimension,'' \emph{arXiv preprint arXiv:1805.11792}, 2018.

\bibitem{zhang2020bayesian}
S.~Zhang, F.~Yang, D.~Zhou, and X.~Zeng, ``Bayesian methods for the yield
  optimization of analog and sram circuits,'' in \emph{2020 25th Asia and South
  Pacific Design Automation Conference (ASP-DAC)}.\hskip 1em plus 0.5em minus
  0.4em\relax IEEE, 2020, pp. 440--445.

\bibitem{bergstra2012random}
J.~Bergstra and Y.~Bengio, ``Random search for hyper-parameter optimization,''
  \emph{The Journal of Machine Learning Research}, vol.~13, no.~1, pp.
  281--305, 2012.

\bibitem{johnson2014nlopt}
S.~G. Johnson, ``The nlopt nonlinear-optimization package,'' 2014.

\bibitem{jones2001scipy}
E.~Jones, T.~Oliphant, P.~Peterson \emph{et~al.}, ``Scipy: Open source
  scientific tools for python,'' 2001.

\end{thebibliography}

\end{document}